\documentclass[10pt,twocolumn,twoside]{IEEEtran}
\usepackage{latexsym,amsmath,amssymb,amsfonts,mathrsfs,amsbsy}
\usepackage[dvips]{graphicx}
\usepackage[final]{epsfig}
\usepackage{cases}
\usepackage{url}
\usepackage{multirow}
\usepackage{array,colortbl}
\usepackage{color}
\usepackage{textpos}
\usepackage{hyperref}
\usepackage{algorithm}
\usepackage{algorithmic}
\usepackage{array}
\usepackage{amssymb}
\usepackage{times}
\usepackage{setspace}
\usepackage{cite}
\usepackage{epstopdf} 
\usepackage{tabularx,ragged2e,booktabs,caption}
\usepackage{mathrsfs}
\usepackage{amsmath}
\usepackage{arydshln}
\begin{document}
\title{Collision Tolerant Packet Scheduling for Underwater Acoustic Localization}

\normalsize

\author{$^\uparrow$Hamid Ramezani*,~\IEEEmembership{Student Member,~IEEE},
		$^\dagger$Fatemeh Fazel,~\IEEEmembership{Member,~IEEE},\\
		$^\dagger$Milica Stojanovic,~\IEEEmembership{Fellow,~IEEE},
        and $^\uparrow$Geert Leus,~\IEEEmembership{Fellow,~IEEE}
\thanks{$^\uparrow$The authors are with the Faculty of Electrical Engineering, Mathematics and Computer Science, Delft University of Technology, 2826 CD Delft, The Netherlands. e-mails: \{h.mashhadiramezani,~g.j.t.leus\}@tudelft.nl.

$^\dagger$The authors are with the Department of Electrical and Computer Engineering, Northeastern University, 02611, MA, USA. e-mails: \{ffazel,millitsa\}@ece.neu.edu.
}
\thanks{* Corresponding author: Hamid Ramezani, phone: (+31)152786280, fax: (+31)152786190, e-mail: h.mashhadiramezani@tudelft.nl.}
\thanks{The research leading to these results has received funding in part from the European Commission FP7-ICT Cognitive Systems, Interaction, and Robotics under the contract \#270180 (NOPTILUS), NSF grant CNS-1212999, and ONR grant N00014-09-1-0700. 
Part of this work is accepted in the Proceeding of IEEE ICC 2014 Workshop on Advances in Network Localization and Navigation (ANLN), 10-14 June 2014, Sydney, Australia \cite{Ramezani2014ICC}.}
}
\markboth{(draft)}{h}
\maketitle
\begin{abstract}
This article considers the joint problem of packet scheduling and self-localization in an underwater acoustic sensor network where sensor nodes are distributed randomly in an operating area. 
In terms of packet scheduling, our goal is to minimize the localization time, and to do so we consider two packet transmission schemes, namely a collision-free scheme (CFS), and a collision-tolerant scheme (CTS). The required localization time is formulated for these schemes, and through analytical results and numerical examples their performances are shown to be generally comparable. However, when the packet duration is short (as is the case for a localization packet), and the operating area is large (above 3km in at least one dimension), the collision-tolerant scheme requires a smaller localization time than the collision-free scheme.
After gathering enough measurements, an iterative Gauss-Newton algorithm is employed by each sensor node for self-localization, and the Cram\'er Rao lower bound is evaluated as a benchmark.
Although CTS consumes more energy for packet transmission, it provides a better localization accuracy. Additionally, in this scheme the anchor nodes work independently of each other, and can operate asynchronously which leads to a simplified implementation.
\end{abstract}
\IEEEpeerreviewmaketitle
\section{Introduction}
\label{sec:introduction}
After the emergence of autonomous underwater vehicles (AUVs) in the 70s, developments in computer systems and networking have been paving a way towards fully autonomous underwater acoustic sensor networks (UASNs) \cite{paull2013auv}, \cite{chatzicristofis2012noptilus}. Modern underwater networks are expected to handle many tasks automatically.
To enable applications such as tsunami monitoring, oil field inspection or shoreline surveillance, the sensor nodes measure various environmental parameters, encode them into data packets, and exchange the packets with other sensor nodes or send them to a fusion center.
The data packets are usually meaningless if they are not labeled with the time and the location of their origin. In this sense, localization is an indispensable task for the network.\\
Due to the challenges of underwater acoustic communications such as low data rates and long propagation delays with variable sound speed \cite{stojanovic2009underwater}, a variety of localization algorithms have been introduced and analyzed in the literature \cite{Han2012Loc} \cite{Erol2011}. These algorithms are relatively different from the ones studied for terrestrial wireless sensor networks (WSNs). For instance, in the terrestrial WSNs, a sensor node can be equipped with a GPS module to determine its location. On the other hand, GPS signals (radio-frequency signals) are highly attenuated underwater, and cannot propagate more than a few meters. Therefore, acoustic signals are usually used for underwater communications. In addition, sensor nodes in WSNs experience low propagation delays in packet exchanging because radio-frequency signals travel almost with light speed. In contrast, acoustic signals propagate very slowly in comparison with light speed, and that introduces long propagation delays between the underwater nodes. 

For an underwater sensor node to determine its location, it can measure the time of flight (ToF) to several anchors with known positions, estimate its distance to them, and then perform multilateration. Other approaches may be employed for self-localization, such as finger-printing \cite{HamidFingerPrinting} or angle of arrival estimation \cite{Kułakowski20101181}. Nevertheless, packet transmissions from anchors are required in all these approaches.

Many factors determine the accuracy of self-localization. Other than noise, the number of anchors, their constellation, relative position of the sensor node \cite{chepuri2013sparsity}, propagation losses and fading also affect the localization accuracy. Some of these parameters can be adjusted to improve the localization accuracy.

Although a great deal of research exists on underwater localization algorithms \cite{paull2013auv}, little work has been done to determine how the anchors should transmit their packets to the sensor nodes. In long base-line (LBL) systems where transponders are fixed on the sea floor, an underwater node interrogates the transponders for round-trip delay estimation \cite{stuart2005acoustic}. 
In the underwater positioning scheme of \cite{cheng2007range}, a master anchor sends a beacon signal periodically, and other anchors transmit their packets in a given order after the reception of the beacon from the previous anchor. 
The localization algorithm in \cite{othman2008gps} addresses the problem of joint node discovery and collaborative localization without the aid of GPS. The algorithm starts with a few anchors as primary seed nodes, and as it progresses, suitable sensor nodes are converted to seed nodes to help in discovering more sensor nodes. The algorithm works by broadcasting command packets which the nodes use for time-of-flight measurements. The authors evaluate the performance of the algorithm in terms of the average network set-up time and coverage. However, physical factors such as packet loss due to fading or shadowing and collisions have not been reviewed, and it is not established whether this algorithm is optimal for localization. 
In reactive localization \cite{watfa2010reactive}, an underwater node initiates the process by transmitting a ``hello'' message to the anchors in its vicinity, and those anchors that receive the message correctly transmit their packets. An existing medium access control (MAC)  protocol may be used for packet exchanging \cite{shahabudeen2013analysis}; however, there is no guarantee that it will perform satisfactorily for the localization task. 
The performance of localization under different MAC protocols is evaluated in \cite{kim2011impact}, where it is shown that a simple carrier sense multiple access (CSMA) protocol performs better than the recently introduced underwater MAC protocols such as T-Lohi \cite{Syed2008a}. 
 
In our previous work, we considered optimal collision-free packet scheduling in a UASN for the localization task in single-channel \cite{Ramezani2013lmac} and multi-channel \cite{Ramezani2013ldmc} scenarios. There, the position information of the anchors is used to minimize the localization time. In spite of the remarkable performance over other algorithms (or MAC protocols), they are highly demanding. Their main drawback is that they require a fusion center which gathers all the position information of the anchors, and decides on the time of packet transmission from each anchor. In addition, they need the anchors to be synchronized and equipped with radio modems in order to exchange information fast. 
In contrast, in this paper we consider packet scheduling algorithms that do not need a fusion center or synchronized anchors. 

We assume a single-hop UASN where anchors are equipped with half-duplex acoustic modems, and can broadcast their packets based on two classes of scheduling: a collision-free scheme (CFS), where the transmitted packets never collide with each other at the receiver, and a collision-tolerant scheme (CTS), where the collision probability is controlled by the packet transmission rate in such a way that each sensor node can receive sufficient error-free packets for self-localization. Our contributions are listed as below.

\begin{itemize}
\item[$\bullet$] Under the conditions of packet loss and collision, the localization time is formulated for each scheme, and its minimum is obtained analytically for a predetermined probability of successful localization for each sensor node.
\item[$\bullet$] An iterative Gauss-Newton self-localization algorithm is introduced for a sensor node which experiences packet loss or collision. Furthermore, it is explained how this algorithm can be used for each packet scheduling scheme. 
\item[$\bullet$] The Cram\'er Rao lower bound (CRB) on localization is derived for each scheme. Other than the distance-dependent signal to noise ratio on each measurement, the effects of packet loss due to fading or shadowing, collisions, and the probability of successful self-localization are included in this derivation.
\end{itemize}

The structure of the paper is as follows. Section~\ref{sec:systemModel} describes the system model, and explains self-localization. The problem of minimizing the localization time in the collision-free and collision-tolerant packet transmission schemes is formulated and analyzed in Section~\ref{sec:collisionFree} and Section~\ref{sec:collisionTolerant}, respectively. The self-localization algorithm is introduced in Section~\ref{sec:LocAlgorithms}. The average energy consumption is analyzed in Section~\ref{sec:energyConsumption}, and Section~\ref{sec:numericalResults} compares the two classes of localization packet scheduling through several numerical examples. Finally, we conclude the paper in Section~\ref{sec:conclusion}, and outline some future work.

\section{System model}
\label{sec:systemModel}
\begin{figure}[b]
\centering
\includegraphics[width=0.45\linewidth]{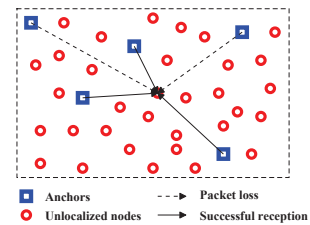}
\caption{\footnotesize{Anchors and sensor nodes are uniformly distributed in a rectangular area.}}
\label{fig:ICC14_Network}
\end{figure}

We consider a UASN consisting of $M$ sensor nodes and $N$ anchors as shown in Fig.~\ref{fig:ICC14_Network}. Each anchor in the network encapsulates information about its ID, its location, time of packet transmission, and a predetermined training sequence for the time of flight estimation. The so-obtained localization packet is broadcast to the network based on a given protocol, e.g., periodically, or upon the reception of a request from a sensor node \cite{Carroll2012ondeman}. The system structure is specified as follows.

\begin{itemize}
  \item[$\bullet$] Anchors and sensor nodes are equipped with half-duplex acoustic modems, i.e., they cannot transmit and receive simultaneously.
  \item[$\bullet$] Sensor nodes are located randomly in an operating area according to some probability density function (pdf). We assume that the distance between a sensor node and an anchor is distributed according to a pdf $g_D(z)$. It is further assumed that the pdf of the distance between the anchors is $f_D(z)$. The pdfs can be estimated from the empirical data gathered during past network operations. 
  \item[$\bullet$] Although the concept of this article can be extended to a multi-hop network, in this work we consider a single-hop network where all the nodes are within the communication range of each other.
  \item[$\bullet$] It is assumed that in the absence of collision, the probability of packet loss due to fading or shadowing between an anchor and a sensor node is $p_l$ which is further assumed to be distance-independent.
\end{itemize}
 
The considered localization algorithms are assumed to be based on ranging, whereby a sensor node determines its distance to several anchors via ToF or round-trip-time (RTT).
Each sensor node can determine its location if it receives at least $K$ different localization packets from $K$ different anchors. The value of $K$ depends on the geometry (2D or 3D), and other factors such as whether depth information of the sensor node is available, and whether the sound speed estimation is required or not. The value of $K$ is usually $3$ for a 2D operating environment with known sound speed and $4$ for a 3D one. In a situation where the underwater nodes are equipped with pressure sensors, three different successful packets would be enough for a 3D localization algorithm \cite{Ramezani2013RangeJ}.

The localization procedure starts either periodically for a predetermined duration (in a synchronized network), or upon receiving a request from a sensor node (in any kind of network, synchronous or asynchronous)
as explained below.
\begin{itemize}
\item[M1)] {\bf Periodic localization}: If all the nodes in the network including anchors and sensor nodes are synchronized with each other, a periodic localization approach may be employed. In this approach, after the arrival of a packet from the $j$-th anchor, the $m$-th sensor node can estimate its distance to that anchor as $\hat{d}_{m,j} = c(\hat{t}^\text{R}_{m,j}-t^\text{T}_j)$ where $c$ is the sound speed, $t^\text{T}_j$ is the time at which the anchor transmits its packet, and $\hat{t}^\text{R}_{m,j}$ is the estimated time at which the sensor node receives this packet. The departure time ${t}^\text{T}_j$ is obtained by decoding the received packet (the anchor inserts this information in the localization packet), and the arrival time $\hat{t}^\text{R}_{m,j}$ can be calculated by e.g., correlating the received signal with the known training sequence. The estimated time of arrival is related to the actual arrival time through $\hat{t}^\text{R}_{m,j} = t^\text{R}_{m,j} + n_{m,j}$, where $n_{m,j}$ is zero-mean Gaussian noise with power $\sigma^2_{m,j}$ which varies with distance and can be modeled as \cite{cardinali2006uwb}
\begin{equation}
\footnotesize
\label{eq:tofNoisePower}
\sigma^2_{m,j} = k_{E}d_{m,j}^{n_0},
\end{equation}
with $d_{m,j}$ the distance between the $j$-th anchor and the $m$-th sensor node, $n_0$ the path-loss exponent (spreading factor), and $k_{E}$ a constant that depends on system parameters (such as signal bandwidth, sampling frequency, channel characteristics, and noise level).

\item[M2)] {\bf On-demand localization}: In this procedure (which can be applied to a synchronous or an asynchronous network) a sensor node initiates the localization process. This is handled by transmitting a high power frequency tone immediately before the request packet. This tone wakes up the anchors from their idle mode, and puts them in the listening mode. The request packet may be used for a more accurate estimation of the arrival time. 
In this paper, we assume that all the anchors have been correctly notified by this frequency tone. After the anchors receive this frequency tone, they reply with their localization packets. In this case,  the time when the request has been received by an anchor and the time at which a localization packet is transmitted have to be included in the localization packet, and this information will be used by the sensor node to estimate its round-trip-time (which is proportional to twice the distance) to the anchor. The round-trip-time can be modeled as 
\begin{equation}
\footnotesize
\hat{t}^\text{RTT}_{m,j} = (t^\text{R}_{m,j}-t^\text{T}_m)-(t^\text{R}_{j,m}-t^\text{T}_j)+n_{j,m}+n_{m,j}.
\end{equation}
Therefore, the estimated distance to anchor $j$ would be

\begin{equation}
\footnotesize
\hat{d}_{m,j}=\frac{1}{2}c\hat{t}^\text{RTT}_{m,j}.
\end{equation}
After the sensor node estimates its location, it broadcasts its position to other sensor nodes. This enables the sensor nodes which have overheard the localization process to estimate their positions without initializing another localization task. 
\end{itemize}

The time it takes for an underwater node to gather at least $K$ different packets from $K$ different anchors is called the localization time. A shorter localization time allows for a more dynamic network, and leads to a better network efficiency in terms of throughput. 
In the next section, we formally define the localization time, and show how it can be minimized for the collision-free and collision-tolerant packet transmission schemes. 

\section{Packet scheduling}
\label{sec:packetDetection}
\subsection{Collision-free packet scheduling}
\label{sec:collisionFree}
\begin{figure}
\centering
\includegraphics[width=0.750\linewidth]{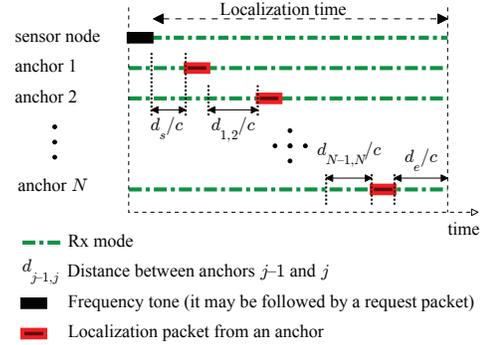}
\caption{\footnotesize{Packet transmission from anchors in the collision-free scheme. Here, each anchor transmits its packets according to its index value (ID number). All links between anchors are assumed to function properly in this figure (there are no missing links).}}
\label{fig:ICCJ_OnDemandRQ_CF}
\end{figure}

Collision-free localization packet transmission is analyzed in \cite{Ramezani2013lmac}, where it is shown that in a fully-connected (single-hop) network, based on a given sequence of the anchors' indices, each anchor has to transmit immediately after receiving the previous anchor's packet. Furthermore, it is shown that there exists an optimal ordering sequence which minimizes the localization time. However, to obtain that sequence, a fusion center is required that knows the positions of all the anchors. In a situation where this information is not available, we may assume that anchors simply transmit in order of their ID numbers as illustrated in Fig.~\ref{fig:ICCJ_OnDemandRQ_CF}. 

\begin{figure}
\centering
\includegraphics[width=0.750\linewidth]{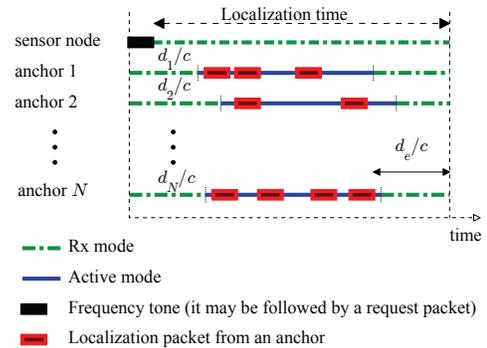}
\caption{\footnotesize{Packet transmission from anchors in the collision-tolerant scheme. Here, each anchor transmit its packets at random according to a Poisson distribution.}}
\label{fig:ICCJ_OnDemandRQ_CT}
\end{figure}

In the event of a packet loss, a subsequent anchor will not know when to transmit. If an anchor does not receive a packet from a previous anchor, it waits for a predefined time (counting from the starting time of the localization process), and then transmits its packet, similarly as introduced in \cite{tan2009enhanced}. With a slight modification of the result from \cite{tan2009enhanced}, the waiting time for the $j$-th anchor could be as short as $t_{j,k}+(j-k)\left(T_p+\frac{D_\text{aa}}{c}\right)$, where $k$ is the index of the anchor whose packet is the last one which has been received by the $j$-th anchor, $t_{j,k}$ is the time at which this packet was received by the $j$-th anchor (counting from the starting time of the localization process), $c$ is the sound speed, $\frac{D_\text{aa}}{c}$ is the maximum propagation delay between two anchors, and $T_p$ is the packet length. The packet length is related to the system bandwidth $B$ (or symbol time $T_s \approx \frac{1}{B}$), number of bits in each symbol $b_s$, number of bits in each packet $b_p$, and guard time $T_g$ as formulated in 
\begin{equation}
\footnotesize
\label{eq:packetLength}
T_p = T_g + \frac{b_p}{b_s}T_s.
\end{equation}
Let us denote by ${\bf u}$ a $(N-1)\times 1$ vector whose element $u_j$ is either 1 or 0, where $u_j= 1$ indicates that there is a packet loss (with probability $p_l$) between anchors $j$ and $j+1$, and $u_j= 0$ represents no packet loss between these anchors. Let us also define ${\bf d}$ as a $(N-1)\times 1$ vector whose $j$-th element, $d_{j,j+1}$, is the distance between the anchors $j$ and $j+1$. With these definitions, the localization time for the collision-free scheme can be formulated as 
\begin{equation}
\footnotesize
\label{eq:locTimeExact}
T_{_\text{CF}} = \frac{1}{c}({\bf 1}_{}-{\bf u})^T{\bf d}+\frac{D_\text{aa}}{c}{\bf u}^T{\bf 1}_{}+NT_p + \frac{d_s}{c} + \frac{d_e}{c},
\end{equation}
where $d_s$ is the distance between the requesting sensor node and anchor node 1 which initiates the localization process, $d_e$ is the distance between the last anchor and the farthest sensor node in the network, and ${\bf 1}$ is a $(N-1)\times 1$ vector whose entries are 1.
Since the anchors are unaware of the position of the farthest sensor node, we set $d_e = D_\text{sa}$, where $D_\text{sa}$ is the distance corresponding to the maximally separated sensor-anchor pair. The addition of this maximum propagation delay ensures that the last transmitted packet will reach the farthest node. Note that in the case of periodic localization $d_s=0$ because no request from a sensor node is required in order to start the localization process.
Given ${\bf u}$, the conditional pdf of $T_{_\text{CF}}$ depends only on the pdfs of the distances between anchor-anchor and anchor-sensor pairs, and can be obtained as 
\begin{equation}
\footnotesize
\label{eq:FTCondOnU}
\begin{aligned}
f_{T_{_\text{CF}}|{\bf u}}(t) &= \\
&{c}\underbrace{f_D(ct) \ast f_D(ct) \ast \hdots \ast f_D(ct)}_{N-1-n_{\bf u} \text{~ times}} \\
                   &\ast g_D(ct) \ast \delta\left(t-NT_p-n_{\bf u}\frac{D_\text{aa}}{c}-\frac{D_\text{sa}}{c}\right),
\end{aligned}
\end{equation}
where $n_{\bf u}$ is the number of lost packets, and $\delta(t)$ is the Dirac function. Using \eqref{eq:FTCondOnU}, the pdf of $T_{_\text{CF}}$ can be obtained as
\begin{equation}
\footnotesize
\label{eq:pdfCFS_LocT}
f_{T_{_\text{CF}}}(t) = \sum_{n=0}^{N-1}{f_{T_{_\text{CF}}|{\bf u}}(t)P(n_{\bf u}=n)}
\end{equation} 
where 
\begin{equation}
\footnotesize
P(n_{\bf u}=k) = \left(\begin{matrix} N-1 \\ k \end{matrix}\right)p_l^k(1-p_l)^{N-1-k}.
\end{equation}
Given the pdf of the collision-free localization time, $f_{T_{_\text{CF}}}(z)$, the minimum collision-free localization time $T_{_\text{CF}}^\text{min}$ for the $N$ anchors to finish transmitting their packets with probability $P_{tt}$ can be obtained by solving
\begin{equation}
\footnotesize
P_{tt} = \int_{0}^{T_{_\text{CF}}^\text{min}}f_{T_{_\text{CF}}}(t)dt.
\end{equation}
Since $d_{j,j+1}$ for $j=1,...,N-1$, $d_s$, and ${\bf u}$ are independent of each other, the average localization time can be obtained as
\begin{equation}
\footnotesize
\begin{aligned}
T_{_\text{CF}}^\text{avg} = & NT_p + (N-1)(1-p_l)\frac{d_\text{avg}}{c} + 
 \\&(N-1)p_l\frac{D_\text{aa}}{c}+\frac{D_\text{sa}}{c} + \frac{\bar{d}_s}{c},
\end{aligned}
\end{equation}
where $d_\text{avg}$ is the average distance between two anchors, and can be formulated as
\begin{equation}
\footnotesize
\label{eq:d_avg}
d_\text{avg} = \int_{0}^{D_\text{aa}}{zf_D(z)dz},
\end{equation}
and $\bar{d}_s$ is the average of $d_s$ which can be either 0 or $d_{s,\text{avg}}$ (periodic or on-demand localization, respectively), with $d_{s,\text{avg}}$ obtained similarly as \eqref{eq:d_avg}.

Under the condition of no packet loss ($p_l=0$), the anchors do not need to wait for the maximum propagation delay, and they can transmit their packets immediately after the complete reception of the previously transmitted packet. As a result, the localization time is the shortest and given by
\begin{equation}
\footnotesize
\label{eq:locTimeBest}
T_{_\text{CF}|_{{\bf u}={\bf 0}}}= NT_p+\frac{1}{c}\sum_{j=1}^{N-1}{d_{j,j+1}}+\frac{d_s}{c}+\frac{D_\text{sa}}{c}.
\end{equation}
For a given $P_{tt}$ the lower bound on the localization time, $T_{_\text{CF}}^\text{min}$, denoted as $T_{_\text{CF}}^\text{low}$, can now be obtained by solving
\begin{equation}
\footnotesize
P_{tt} = \int_{0}^{T_{_\text{CF}}^\text{low}}f_{T_{_\text{CF}|_{{\bf u}={\bf 0}}}}(z)dz.
\end{equation}

In contrast, in the worst case, all the packets between anchors are lost, and the requesting sensor node is located at its farthest distance to the initiating anchor. This case yields the longest localization time given by
\begin{equation}
\footnotesize
T_{_\text{CF}}^\text{upp}= NT_p+(N-1)\frac{D_\text{aa}}{c}+\frac{D_\text{sa}}{c}+\frac{D_\text{sa}}{c},
\end{equation}
which is equivalent to packet transmission based on time division multiple access (TDMA) with time-slot duration $T_p+\frac{D}{c}$ (assuming $D=D_\text{sa}=D_\text{aa}$).


Another figure of merit is the probability with which a node can localize itself. If this probability is required to be above a design value $P_{ss}$, the necessary number of anchors is determined as the smallest $N$ for which  
\begin{equation}
\footnotesize
\label{eq:CF_anchorNum}
P^\text{loc}_{_\text{CF}}=\sum_{k=K}^{N}{\left(\begin{matrix} 
N \\ k
\end{matrix}\right)p_{_\text{CF}}^k(1-p_{_\text{CF}})^{N-k}} \geq P_{ss}
\end{equation}
where $p_{_\text{CF}}$ is the probability that a transmitted packet reaches a sensor node correctly, and it can be calculated as 
\begin{equation}
\footnotesize
p_{_\text{CF}} = (1-p_l)\int_{\gamma_0N_0B}^{\infty}{f_{X_0}(x)dx}, 
\end{equation}
where $N_0B$ is the noise power, $\gamma_0$ is the minimum SNR at which a received packet can be detected at the receiver, and $f_{X_0}(x)$ is the pdf of the received signal power which will be derived in the next subsection. Note that in one-hop communications, the transmission power will be set in such a way that for any distance, in the collision-free scheme, the SNR is greater than $\gamma_0$, i.e., $p_{_\text{CF}} = (1-p_l)$.

It is worth mentioning that instead of increasing the number of anchors, in a mobile scenario one can repeat packet transmissions from $K$ anchors multiple times. That would change \eqref{eq:locTimeExact} and the pdf of the localization time \eqref{eq:pdfCFS_LocT} to some extent; however, this approach is not considered in the present analysis.

\subsection{Collision-tolerant packet scheduling}
\label{sec:collisionTolerant}
To avoid the need for coordination among anchor nodes, in a collision-tolerant packet scheduling, anchors work independently of each other. During a localization period or upon receiving a request from a sensor node, they transmit randomly, e.g. according to a Poisson distribution with an average transmission rate of $\lambda$ packets per second. Packets transmitted from different anchors may now collide at a sensor node, and the question arises as to what is the probability of successful reception. This problem is a mirror image of the one investigated in \cite{fazel2013random} where sensor nodes transmit their packets to a common fusion center. Unlike \cite{fazel2013random} however, where the sensors know their location, and power control fully compensates for the known path-loss, path-loss is not known in the present scenario, and there is no power control. The average received signal strength is thus different for different links (this signal strength, along with a given fading model, determines the probability of packet loss). In this regard, the signal received at the $m$-th sensor node from the $j$-th anchor is 
\begin{equation}
\footnotesize
v_{m,j}(t)= c_{m,j}v_j(t)+i_m(t)+w_m(t),
\end{equation}
where $v_j(t)$ is the signal transmitted from the $j$-th anchor, $c_{m,j}$ is the channel gain, $w_m(t)$ is the additive white Gaussian noise with power $N_0B$, and $i_m(t)$ is the interference caused by other anchors whose packets overlap with the desired packet,
\begin{equation}
\footnotesize
i_m(t)= \sum_{k\neq j}^{ }{c_{m,k}v_k(t-\tau_k)},
\end{equation}
where $\tau_k$ is the difference in the arrival times of the interfering signals w.r.t. the desired signal, and it is modeled as an exponentially distributed random variable. 
The SNR at the receiver depends on the interference level, and is given by
\begin{equation}
\footnotesize
\gamma = \frac{X_0}{I_0+N_0B},
\end{equation}
where $X_0 = |c_{m,j}|^2P_0$ is the power of the signal of interest with $P_0$ the anchor's transmit power, and where $I_0$ is the total interference power which can be  expressed as
\begin{equation}
\footnotesize
I_0=\sum_{i=1}^{q}{|c_{m,k_i}|^2P_0} 
\end{equation}
with $q$ the number of interferers, and $k_i$ the index of the $i$-th interferer. Using a simple path-loss model we can formulate the attenuation of the signal power as
\begin{equation}
\footnotesize
\label{eq:signalPower}
|c_{m,j}|^2=\alpha_0\left(d_0/d_{m,j}\right)^{n_0},
\end{equation}
where $\alpha_0$ is a constant and $d_0$ is the reference distance. Using \eqref{eq:signalPower}, the pdf of the received signal power of the desired signal is
\begin{equation}
\footnotesize
f_{X_0}(x) = \frac{d_0}{n_0}\left(P_0\alpha_0\right)^\frac{1}{n_0}\left(\frac{1}{x}\right)^{\frac{1}{n_0}+1}g_d\left(d_0(\frac{P_0\alpha_0}{x})^\frac{1}{n_0}\right),
\end{equation}
and the pdf of the interference can be obtained as
\begin{equation}
\footnotesize
f_{I_0}(x) = \underbrace{f_{X_0}(x) \ast f_{X_0}(x) \ast \hdots \ast f_{X_0}(x)}_{q \text{~ times}}.
\end{equation}
The probability that a packet is received correctly by a sensor node is then \cite{fazel2013random}
\begin{equation}
\footnotesize
\label{eq:collisionFreePs}
p_s = (1-p_l)\sum\limits_{q=0}^{N-1}{P(q)p_{s|q}}, 
\end{equation}
where $P(q)=\frac{(2N\lambda T_p)^q}{q!}e^{-2N\lambda Tp}$ is the probability that $q$ packets interfere with the desired packet, and $p_{s|q}$ is the probability that the desired packet ``survives'' under this condition,
\begin{equation}
\footnotesize
\begin{split}
\begin{aligned}
& p_{s|q} = \\ 
       &\begin{cases}
                  \int_{\gamma_0N_0B}^{\infty}{f_{X_0}(x)dx} & \text{}q=0 \\ 
                  \int_{\gamma_0}^{\infty}{\int_{N_0B}^{\infty}{f_{X_0}(\gamma w)f_I(w-N_0B)}wdwd\gamma} & \text{}q\geq 1
       \end{cases}
\end{aligned}
\end{split}
\end{equation}
where $w=I_{0}+N_0B$. 

In addition, it should be noted that redundant successfully received packets from an anchor are not useful for localization. However, they may be used to reduce the effects of noise on the range estimation (see Section \ref{sec:LocAlgorithms}), or in mobile scenarios where the anchors are moving they can be used for range tracking \cite{ramezani2011target}. However, we will not consider the mobile case in this paper.

The probability of receiving a useful packet from an anchor during transmission time $T_{_\text{T}}$ can now be approximated by \cite{fazel2013random}
\begin{equation}
\footnotesize
\label{eq:collisionTolPg}
p_{_\text{CT}} = 1-e^{-p_s\lambda T_{_\text{T}}},
\end{equation}
and the probability that a sensor node accomplishes self-localization during $T_{_\text{T}}$ seconds using $N$ anchors can be obtained as
\begin{equation}
\footnotesize
\label{eq:collisionFreePss}
P^\text{loc}_{_\text{CT}} = \sum_{k=K}^{N}{\left(\begin{matrix}
 N \\ k
\end{matrix}\right)p_{_\text{CT}}^k(1-p_{_\text{CT}})^{N-k}},
\end{equation}
which is equivalent to the probability that a node receives at least $K$ different localization packets during $T_{_\text{T}}$ seconds. 

It can be shown that $P^\text{loc}_{_\text{CT}}$ is an increasing function of $T_{_\text{T}}$ (see Appendix \ref{app:p_loc_T}), and as a result for any value of $p_s\lambda \neq 0$, there is a $T_{_\text{T}}$ that leads to the probability of self-localization equal or greater than $P_{ss}$. The minimum value for the required $T_{_\text{T}}$ can be obtained at a point where $p_s\lambda$ is maximum ($\lambda_\text{opt}$). It can be proven that the lower bound of $\lambda_\text{opt}$ is $\lambda^\text{low}_\text{opt} = \frac{1}{2NT_p}$, and its upper bound is $\frac{N+1}{2NT_p}$ (see Appendix \ref{app:ps_lambda}). 

Given the number of anchors $N$, and a desired probability of successful self-localization $P_{ss}$, one can determine $p_{_\text{CT}}$ from \eqref{eq:collisionFreePss}, and $\lambda$ and the minimum localization time jointly from \eqref{eq:collisionFreePs} and \eqref{eq:collisionTolPg}. Similarly to the collision-free scheme, we then add the time of request $\frac{d_s}{c}$, and the maximum propagation delay between anchor-sensor pair  $\frac{D_\text{sa}}{c}$ to the (minimum) $T_{_\text{T}}$ that is obtained from \eqref{eq:collisionFreePs} and \eqref{eq:collisionTolPg}. This value is then considered as the (minimum) localization time ($T_{_\text{CT}}^\text{min}$) $T_{_\text{CT}}$, for the collision-tolerant scheme.

\section{Self-Localization process}
\label{sec:LocAlgorithms}
As explained before, a sensor node requires at least $K$ distinct packets (or time-of-flight measurements) in order to determine its location. However, it may receive more than $K$ different packets as well as some replicas, i.e, $q_j$ packets from anchor $j$, where $j=1,...,N$. In this case, it uses all of this information for self-localization. Note that in the collision-free scheme, $q_j$ is either zero or one; however, in the collision-tolerant scheme $q_j$ can be more than 1. Packets received from the $j$-th anchor can be used to estimate the sensor node's distance to that anchor, and the redundant packets would add diversity (or reduce measurement noise) for this estimate. In the next two subsections, we show how all of the correctly received packets can be used in a localization algorithm, and how the CRB of the location estimate can be obtained for the proposed scheduling schemes.

\subsection{Localization algorithm}
After the anchors transmit their localization packets, each sensor node has $\mathcal{Q}$ measurements. Each measurement is contaminated by noise whose power is related to the distance between the sensor and the anchor from which the measurement has been obtained. The $l$-th measurement obtained from the $j$-th anchor is related to the sensor's position, ${\bf x}$, as (sensor node index is omitted for simplicity)
\begin{equation}
\footnotesize
\hat{t}_l=f({\bf x})+n_l,
\end{equation}
where $n_l$ is the noise and $f({\bf x})$ is
\begin{equation}
\footnotesize
f({\bf x}) = \frac{1}{c}\left\|{\bf x}-{\bf x}_j\right\|_2 
\end{equation}
where ${\bf x}_j$ is the $j$-th anchor's position. Stacking all the measurements gives us a $\mathcal{Q} \times 1$ vector $\hat{\bf t}$. The number of measurements can be formulated as
\begin{equation}
\footnotesize
\mathcal{Q} = \sum_{j=1}^{N}{q_j},
\end{equation}
where $q_j$ is the number of measurements which are obtained correctly from the $j$-th anchor. 
In CFS, $q_j$ is a Bernoulli random variable with success probability $p_{_\text{CF}}$, in CTS $q_j$ is a Poisson random variable with distribution
\begin{equation}
\footnotesize
P_j^n = P(q_j = n) = \frac{(p_s\lambda T_{_\text{T}})^n}{n!}{e^{-p_s\lambda T_{_\text{T}}}}.
\end{equation}

Since the measurement errors are independent of each other, the maximum likelihood solution for ${\bf x}$ is given by
\begin{equation}
\footnotesize
\label{eq:LocOptFunc}
\hat{\bf x} = \arg \min_{\bf x}\left\|\hat{\bf t}-{\bf f}({\bf x})\right\|_2,
\end{equation}
which can be calculated using a method such as the Gauss-Newton algorithm specified in Algorithm~\ref{alg:GaussNewtonAlg}. 
\begin{algorithm}
\caption{Gauss-Newton Algorithm}
\label{alg:GaussNewtonAlg}
\begin{algorithmic}
\STATE Start with an initial location guess. 
\STATE Set $i=1$ and $E=\infty$.
\WHILE{$i \leq I$ and $E \geq \epsilon$ }
\STATE Next state: 
\STATE ~~~~${\bf x}^{(i+1)}= {\bf x}^{(i)} - $
\STATE ~~~~ $\eta\left(\nabla{\bf f}({\bf x}^{(i)})^T\nabla{\bf f}({\bf x}^{(i)})\right)^{-1}\nabla{\bf f}({\bf x}^{(i)})^T\left({\bf f}({\bf x}^{(i)})-\hat{{\bf t}}\right)$
\STATE ~~~~$E = ||{\bf x}^{(i+1)}-{\bf x}^{(i)}||$
\STATE ~~~~$i = i + 1$
\ENDWHILE
\STATE $\hat{{\bf x}} = {\bf x}^{(i)}$ 
\end{algorithmic}
\end{algorithm}
In this algorithm, $\eta$ controls the convergence speed, $\nabla{\bf f}({\bf x}^{(i)})=\left[\frac{\partial f_1}{\partial {\bf x}},\frac{\partial f_2}{\partial {\bf x}},\hdots,\frac{\partial f_\mathcal{Q}}{\partial {\bf x}}\right]^T_{{\bf x}={\bf x}^{(i)}}$ represents the gradient of the vector ${\bf f}$ w.r.t. the variable ${\bf x}$ at ${\bf x}^{(i)}$, ${\bf x}^{(i)}$ is the estimate in the $i$-th iteration, and $\frac{\partial f_l}{\partial {\bf x}} = \left[\frac{\partial f_l}{\partial x},\frac{\partial f_l}{\partial y},\frac{\partial f_l}{\partial z}\right]^T$ where $l=1,\ldots,\mathcal{Q}$. Here, $I$ and $\epsilon$ are the user-defined limits on the stopping criterion that determines when the algorithm exits the loop. The initial guess is also an important factor for the algorithm. One may obtain the initial guess through geometrical properties of a triangulation, similarly as explained in \cite{jamali2012cooperative}.

\subsection{Cram\'er-Rao bound}
The Cram\'er-Rao bound is a lower bound on the variance of any unbiased estimator of a deterministic parameter. In this subsection, we derive the CRB for the location estimate of a sensor node. 

In order to find the CRB, the Fisher information matrix (FIM) has to be calculated. The Fisher information is a measure of information that an observable random variable $\hat{\bf t}$ carries about an unknown parameter ${\bf x}$ upon which the pdf of $\hat{\bf t}$ depends. The elements of the FIM are defined as
\begin{equation}
\footnotesize
{\bf I}({\bf x})_{i,j} = -\mathbb{E}\left[\frac{\partial^2 }{\partial x_i \partial x_j}\log{{h}(\hat{\bf t};{\bf x})}|{\bf x}\right]
\end{equation}
where ${\bf x}$ is the location of the sensor node,  ${ h}(\hat{\bf t};{\bf x})$ is the pdf of the measurements parametrized by the value of ${\bf x}$, and the expected value is over the cases where the sensor is localizable.

In a situation where the measurements (ToFs or RTTs between a sensor node and the anchors) are contaminated with Gaussian noise (whose power is related to the SNR or equivalently to the mutual distance between a sensor-anchor pair), the elements of the FIM can be formulated as

\begin{equation}
\footnotesize
\begin{aligned}
&{\bf I}({\bf x})_{i,j} = \frac{1}{P^\text{loc}} \underset{\text{s.t. } \{q_1,...,q_N\} \text{ enable self-localization}}{\sum_{{q_N=0}}^{Q_N} ~~~~\hdots~~~~\sum_{q_2 = 0}^{Q_2}~\sum_{q_1 = 0}^{Q_1}}\\
&\left\{\frac{\partial{\bf f}}{\partial{x_i}}^T {\bf R}_w^{-1}\frac{\partial{\bf f}}{\partial{x_j}}\right. + 
{\left.\frac{1}{2}\text{tr}\left[{\bf R}_w^{-1}\frac{\partial{\bf R}_w}{\partial{x_i}}{\bf R}_w^{-1} \frac{\partial{\bf R}_w}{\partial{x_j}}\right]\right\}\Pi_{j=1}^{N}P^{q_j}_{j}}
\label{eq:FIMeq}
\end{aligned}
\vspace{0.2cm}
\end{equation}
where $P^\text{loc}$ is the localization probability, $Q_i=1$ for CFS, and $\infty$ for CTS, ${\bf R}_w$ is the $\mathcal{Q} \times \mathcal{Q}$ noise covariance matrix
\begin{equation}
\footnotesize
\label{eq:FCOV_EQ}
\frac{\partial{\bf R}_w}{\partial{x_i}} = \text{diag}\left(\frac{\partial{[{\bf R}_w]}_{11}}{\partial{x_i}}, \frac{\partial{[{\bf R}_w]}_{22}}{\partial{x_i}},...,\frac{\partial{[{\bf R}_w]}_{\mathcal{Q} \mathcal{Q}}}{\partial{x_i}} \right),
\end{equation}
and 
\begin{equation}
\footnotesize
\frac{\partial{\bf f}}{\partial{x_i}} = \left[\frac{\partial{f_1}}{\partial{x_i}},\frac{\partial{f_2}}{\partial{x_i}},...,\frac{\partial{f_{\mathcal{Q}}}}{\partial{x_i}}\right]^T,
\end{equation}
with $f_i$ a ToF (or RTT) measurement.

Once the FIM has been computed, the lower bound on the variance of the estimation error can be expressed as $\text{CRB}=\sum_{i=1}^3{\text{CRB}_{x_i}}$
where $\text{CRB}_{x_i}$ is the variance of the estimation error in the $i$-th
variable and it is defined as
\begin{equation}
\footnotesize
\text{CRB}_{x_i} = \left[I^{-1}({\bf x})\right]_{ii}.
\end{equation}

Note that the CRB is meaningful if the node is localizable ($\frac{1}{P^\text{loc}}$ in \eqref{eq:FIMeq}), meaning that a sensor node has at least $K$ different measurements. Hence, only $\sum_{k=K}^{N}\footnotesize{\left(\begin{matrix}N \\k \end{matrix}\right)}$ possible states have to be considered in order to calculate \eqref{eq:FIMeq} for collision-free scheduling, while the number of states is countless for collision-tolerant scheduling. Nonetheless, it can be shown that the number of possible states in CTS can be dropped to that of CFS (see Appendix \ref{app:CRB_CTS}).

\section{Energy consumption}
\label{sec:energyConsumption}
In this section, we consider the average energy consumed during localization. In CFS, the receiver of anchor $j$ is on for $t_j$ seconds. If the power consumption in listening mode is $P_\text{L}$, and in transmitting mode $P_\text{0}$, the total energy consumption in CFS can be formulated as
\begin{equation}
\footnotesize
E_{_\text{CF}} = NT_pP_\text{0}+\sum_{j=1}^{N}{t_jP_\text{L}},
\end{equation}
where the processing energy consumption has been ignored.
The receiving time is a random variable and can be formulated as
\begin{equation}
\footnotesize
t_j = \frac{1}{c}({\bf 1}_j-{\bf u}_j)^T{\bf d}_j + \frac{D_\text{aa}}{c}{\bf u}_j{\bf 1}_j,\text{ for }j={2,...,N}
\end{equation}   
where ${\bf u}_j$ (${\bf 1}_j$) is a $(j-1)\times 1$ vector whose elements are the first $j-1$ elements of ${\bf u}$ ({\bf 1}). Note that in our localization procedures $t_1 = 0$, because the first anchor does not listen to the channel in periodic localization (M1), and the time that it receives a request in on-demand localization (M2) is negligible. The average time that the receiver of each anchor is on can be calculated as 
\begin{equation}
\footnotesize
\label{eq:t_avg_CF}
t^{\text{avg}}_j = \frac{j-1}{c}[ (1-p_l)d_\text{avg} + p_l D_\text{aa}], 
\end{equation}   
which results in 
\begin{equation}
\footnotesize
\label{eq:energy_CF}
\begin{aligned}
E^\text{avg}_{_\text{CF}} = & NT_pP_\text{0}+\\
& P_\text{L}\left[(1-p_l)\frac{d_\text{avg}}{c} + p_l \frac{D_\text{aa}}{c}\right]\frac{N(N-1)}{2},
\end{aligned}
\end{equation}
where $E^\text{avg}_{_\text{CF}}$ is the average energy consumption by CFS during each localization procedure. As is clear from $\eqref{eq:t_avg_CF}$, an anchor with a higher index value consumes more energy in comparison with the one that has a lower index. To overcome this problem, anchors can swap indices in each localization procedure.

In CTS, the anchors do not need to listen to the channel and they only transmit at an average rate of $\lambda$ packets per second. The average energy consumption is thus
\begin{equation}
\footnotesize
\label{eq:energy_CT}
E^\text{avg}_{_\text{CT}} = \lambda T_\text{T}NT_pP_\text{0}.
\end{equation}
For small ratios $\frac{P_\text{L}}{P_\text{0}}$, the average energy consumption of CTS is always greater than that of CFS. However, as $\lambda$ gets smaller (or equivalently $T_{_\text{CT}}$ get larger), the energy consumption by CTS reduces. 
\section{Numerical Results}
\label{sec:numericalResults}
To illustrate the results, a two-dimensional rectangular-shape operating area with length $D_x$ and width $D_y$ is considered with uniformly distributed anchors and sensors. There is no difference in how the anchors and sensor nodes are distributed, and therefore we have $f_D(d)=g_D(d)$ which can be obtained as (see Appendix \ref{app:pdf_d})
\begin{align}
\footnotesize
\label{eq:distPdf}
\footnotesize
& f_{D}(d)= \\ \nonumber 
& \frac{2d}{D_x^2D_y^2}\left[ d^2(\sin^2\theta_e-\sin^2\theta_s) + 2D_xD_y(\theta_e-\theta_s) \right. \\
&          + \left. 2D_xd(\cos\theta_e-\cos\theta_s) - 2D_yd(\sin\theta_e-\sin\theta_s) \right]  \nonumber    
\end{align}
where $\theta_s$ and $\theta_e$ are related to $d$ as given in Table~\ref{tab:thetas_theta_e}.
\begin{table}[h]
\footnotesize
\caption{\footnotesize{Values of $\theta_s$ and $\theta_e$ based on distance $d$.}}
\begin{center}
\begin{tabular}{|c|c|c|}
\hline  distance                            & $\theta_s$                  & $\theta_e$      \\ 
\hline  $0 \leq d \leq D_y$                 & $0$                         & $\frac{\pi}{2}$ \\ 
\hline  $D_y\leq d \leq D_x $               & $0$                         & $\sin^{-1}\frac{D_y}{d}$ \\ 
\hline  $D_y\leq d \leq \sqrt{D_x^2+D_y^2}$ & $\cos^{-1}\frac{D_x}{d}$    & $\sin^{-1}\frac{D_y}{d}$ \\ 
\hline 
\end{tabular}
\label{tab:thetas_theta_e}
\end{center} 
\end{table}

The parameter values for the numerical results are listed in Table~\ref{tab:env}, and for all numerical results, we use these values unless otherwise stated.

\begin{table}
\footnotesize
\centering
\caption{\footnotesize{\footnotesize{Simulation parameters. Note that, in this table some parameters such as $N$, $D_\text{aa}$, $T_g$, etc. are related to other parameters, e.g., $N$ depends on the values of the $p_l$, and $P_{ss}$. }}}
\begin{tabular}{|l||c|c|c|}
\hline
Description                                            & Parameter        & Value      & Unit  \\
\hline \hline
Number of anchor nodes                                 & $N$              & $5$         &  -   \\ 
Number of sensor nodes                                 & $M$              & $100$       &  -   \\ 
Sound speed                                            & $c$              & $1500$      & m/s  \\ 
Number of required different packets                   & $K$              & $3$         &  -   \\ 
Area size in $x$-axis                                  & $D_x$            & $3c$        &  m   \\ 
Area size in $y$-axis                                  & $D_y$            & $3c$        &  m   \\
maximum anchor-anchor distance                         & $D_\text{aa}$    &$3c\sqrt{2}$ &  m   \\
maximum anchor-sensor distance                         & $D_\text{sa}$    &$3c\sqrt{2}$ &  m   \\
Guard time for localization packet                     & $T_g$            & $50$        &  ms  \\ 
Number of bits per sample                              & $b_s$            & $2$         &  -   \\
Number of bits per packet                              & $b_p$            & $200$       &  -   \\
System bandwidth                                       & $B$              & $2$         & kHz  \\ 
Localization packet length                             & $T_p$            & $100$       & ms   \\
Packet loss probability                                & $p_l$            & $0.1$       & -    \\
Noise power                                            & $N_0B$           & $-47.5$     & dB   \\
ToF noise power coefficient                            & $k_E$            & $10^{-8}$   &     \\
Transmit power                                         & $P_0$            & $15$        & w    \\
Reference distance                                     & $d_0$            & $1$         & m    \\
Power coefficient                                      & $\alpha_0$       & $1$         & m    \\
Path-loss exponent                                     & $n_0$            & $1.4$       & -    \\
Required SNR for packet detection                      &$\gamma_0$        & $6$         & dB   \\
Request packet arrival delay                           &$d_s/c$           & $0$         & s    \\
Required probability of successful                     &                  &             &      \\
~~localization									 	   &$P_{ss}$		  & $0.99$      & -     \\
Required probability that all packet are               &                  &             &      \\ 
~~transmitted before $T^\text{min}_{_\text{CF}}$ in CFS& $P_{tt}$         & $0.90$      & -    \\ 
\hline 
\end{tabular}
\label{tab:env}
\end{table}

The number of bits in each packet is set to $b_p = 200$ which is sufficient for the position information of each anchor, time of transmission, (arrival time of the request packet), and the training sequence. Assuming QPSK modulation ($b_s=2$), guard time $T_g=50$ms, and a bandwidth of $B=2$kHz the localization packet length is $T_p = 100$ms (see \eqref{eq:packetLength}). In addition, $k_E$ is set to $10^{-8}$ for the sake of simulation. In theory it can acquire much smaller values. 

\begin{figure}[b]
\centering
\includegraphics[width=0.7\linewidth]{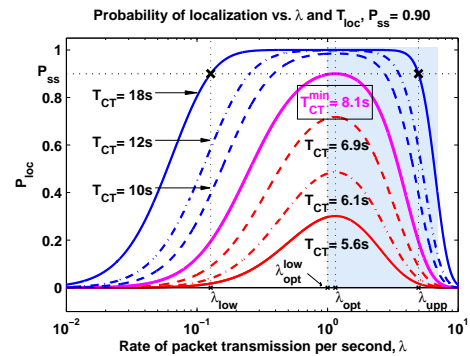}
\caption{\footnotesize{Probability of successful localization for different values of $\lambda$ and $T_{_{\text{CT}}}$.}}
\label{fig:ICC14_PssLambda}
\end{figure}

Fig.~\ref{fig:ICC14_PssLambda} shows the probability of successful self-localization in the collision-tolerant scheme as a function of $\lambda$ and the indicated value for $T_{_{\text{CT}}}$. It can be observed that there is an optimal value of $\lambda$ (denoted by $\lambda_\text{opt}$) which corresponds to the minimal value of $T_{_{\text{CT}}}$ ($T^{\text{min}}_{_{\text{CT}}}$) which satisfies $P^\text{loc}_{_\text{CT}} \geq P_{ss}$. The highlighted area in Fig.~\ref{fig:ICC14_PssLambda} shows the predicted region (obtained in Appendix \ref{app:ps_lambda}) where $\lambda_\text{opt}$ is. As it can be seen, $\lambda_\text{opt}$ is close to $\lambda^\text{low}_\text{opt}$, and it gets closer to this value as $P_{s|q>0}$ gets smaller.
In addition, for the values of $T_{_\text{CT}}$ greater than $T^\text{min}_{_\text{CT}}$, a range of values for $\lambda \in [\lambda_\text{low},\lambda_\text{upp}]$ can attain the desired probability of self-localization. In this case, the lowest value for $\lambda$ should be selected to minimize the transmission energy consumption.

\begin{figure}
\centering
\includegraphics[width=0.7\linewidth]{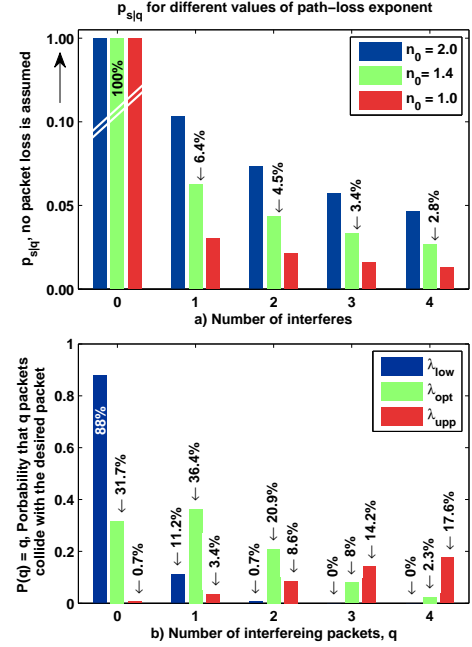}
\caption{\footnotesize{a) Probability of successful packet reception versus different number of interferers.
b) Probability that $q$ interferers collide with the desired packet. For this figure, $\lambda_\text{low}$, $\lambda_\text{opt}$ and $\lambda_\text{upp}$ are chosen from Fig.~\ref{fig:ICC14_PssLambda}.}}
\label{fig:ICC14_PsnPn}
\end{figure}

Fig.~\ref{fig:ICC14_PsnPn} shows the probability of correct packet reception versus the number of interferers (the effect of packet loss due to fading is not included in the figure, and the desired $P_{ss}$ is set to $0.90$ in this example) for different values of the path-loss exponent $n_0$. As it was mentioned before, when there is no interference, the probability of packet reception is 1. Yet, when there is an interferer, the chance of correct reception of a packet becomes very small ($0.06$ for $n_0=1.4$), and as the number of interferers grows, it gets smaller. 

The probability that two or more packets overlap with each other is also depicted in part (b) of this figure for the three values of $\lambda$ shown in Fig.~\ref{fig:ICC14_PssLambda}. It can be seen that as the value of $\lambda$ is reduced from $\lambda_\text{opt}$ (which is equivalent to a larger $T_{_\text{CT}}$), the probability of collision gets smaller. This increases the chance of correct packet reception, and reduces the energy consumption as explained in Section~\ref{sec:energyConsumption}. In addition, it can be observed that although using $\lambda_\text{upp}$ results in the same performance as $\lambda_\text{low}$, it relies on the packets that have survived collisions, which is not energy-efficient in practical situations neither for anchors (required energy for multiple packet transmissions) nor for sensor nodes (processing energy needed  for packet detection).

\begin{figure}
\centering
\includegraphics[width=0.7\linewidth]{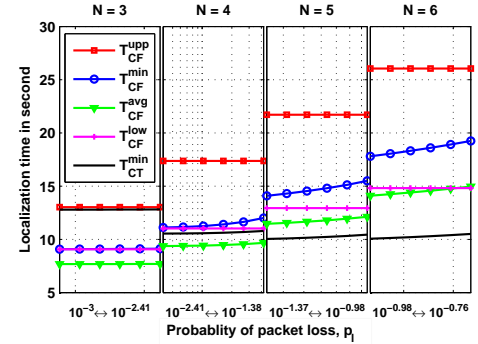}
\caption{\footnotesize{Effect of link-loss probability on the minimum required time for localization. The greater the value of $p_l$ is, the more anchors are required in the collision-free protocols.}}
\label{fig:ICC14_LinkLossProb}
\end{figure}

Fig.~\ref{fig:ICC14_LinkLossProb} shows the minimum required time for localization versus the probability of packet loss. Packet loss is a phenomenon that is common in underwater acoustic systems because of many reasons such as location-dependent fading, shadowing, noise, and so on. As $p_l$ increases, more anchors are required for collision-free localization. In Fig.~\ref{fig:ICC14_LinkLossProb}, for a given $p_{l}$, the number of anchors $N$ is calculated using \eqref{eq:CF_anchorNum}, which is then used to calculate the minimum required time for the collision-free and collision-tolerant localization. Each increase in $T^\text{upp}_{_\text{CF}}$ in CFS indicates that the number of anchors has been increased by one. 
We also note that for a given number of anchors, the lower and upper bounds of the collision-free algorithm are constant over a range of $p_l$ values because they are not affected by that; however, the actual performance of both schemes becomes worse as $p_l$ gets larger. Still, the collision-tolerant approach performs better for a wide range of $p_l$, and as the number of anchors increases its performance slightly changes for high values of $N$. Therefore, it can be used in a system with limited anchors, and can be implemented in practice with low computational complexity since the anchors work independently of each other.

Many factors such as noise power or packet length are directly dependent on the operating frequency and system bandwidth. Assuming single-hop communication among the sensor nodes, an optimum frequency band exists for a given operating area. As the size of the operating area increases, a lower operating frequency (with less bandwidth) is used to compensate for the increased attenuation. Furthermore, as the distance increases the amount of available bandwidth for the optimum operating frequency also gets smaller \cite{Stojanovic:2007:RCD:1347364.1347373}.
As it was mentioned before, the localization packet is usually short in terms of the number of bits, but its duration (in seconds) still depends on the system bandwidth. In this part, we investigate the effect of packet length (or equivalently system bandwidth) on the localization time.

\begin{figure}[]
\centering
\includegraphics[width=0.7\linewidth]{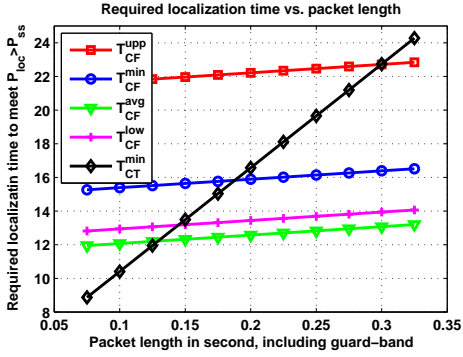}
\caption{\footnotesize{Effect of packet length on the minimum required time for localization.}}
\label{fig:ICC14_Tp}
\end{figure}

\begin{figure}[]
\centering
\includegraphics[width=0.7\linewidth]{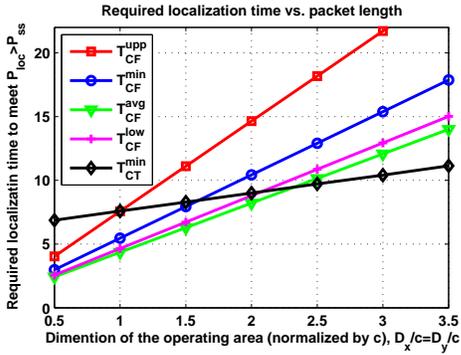}
\caption{\footnotesize{Effect of the operating area size on the required localization time.}}
\label{fig:ICC14_DxDy}
\end{figure}

As it is shown in Fig.~\ref{fig:ICC14_Tp}, the length of the localization packet plays a significant role in the collision-tolerant algorithm. The minimum localization time grows almost linearly w.r.t. $T_p$ in all cases; however, the rate of growth is much higher for the collision-tolerant system than for the collision-free one.
At the same time, as shown in Fig.~\ref{fig:ICC14_DxDy}, the size of the operating area has a major influence on the performance of the CFS, while that of the CTS does not change very much. It can be deduced that in a network where the ratio of the packet length to the maximum propagation delay is low, the collision-tolerant algorithm outperforms the collision-free one.

\begin{figure}[h]
\centering
\includegraphics[width=0.7\linewidth]{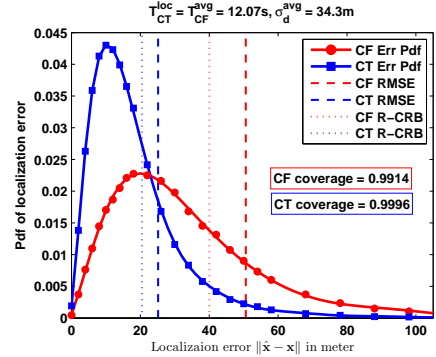}
\caption{\footnotesize{Probability distribution of the localization error, and its corresponding CRB for CTS and CFS.}}
\label{fig:ICC14_ErrorPdfLoc}
\end{figure}

The localization accuracy is related to the noise level at which a ToF measurement is taken, and to the anchors' constellation. If a sensor node in a 2D operating system receives packets from the anchors which are (approximately) located on a line, the sensor node is unable to localize itself (or it experiences a large error). To evaluate the localization accuracy of each algorithm, we considered $M=50$ sensor nodes, and run a Monte Carlo simulation ($10^3$ runs) to extract the results. The number of iterations in Algorithm~1 is set to $I=50$, and the convergence rate is $\eta = \frac{1}{5}$. The $T_{_\text{CF}}$ was set equal to the average localization time of CFS. In this special case where $T^\text{min}_{_\text{CF}}$ is lower than $T^\text{avg}_{_\text{CT}}$, the successful localization probability ($P^\text{loc}$) of CTS would be better than that of CFS. 
The probability distribution of the localization error $\|\hat{{\bf x}}-{\bf x}\|$ is illustrated in Fig.~\ref{fig:ICC14_ErrorPdfLoc} for both schemes. In this figure, the root mean square error (RMSE), and root CRB (R-CRB) are also shown with the dashed and dash-dotted lines, respectively. It can be observed that in CTS the pdf is concentrated at lower values of the localization error in comparison with CFS, because each sensor in CTS has a chance of receiving multiple copies of the same packet, and that reduces the range estimation error.  

\begin{figure}[h]
\centering
\includegraphics[width=0.7\linewidth]{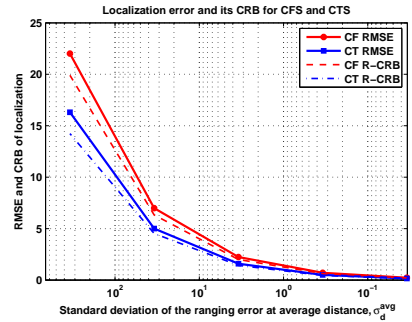}
\caption{\footnotesize{CRB of the localization estimate for each packet scheduling scheme versus $\sigma^\text{avg}_d$.}}
\label{fig:ICC14J_CRB}
\end{figure}

Measurement noise plays a major role in the localization accuracy. For a fixed signal bandwidth, the accuracy of range estimation is only a function of the SNR. Since the distance between the nodes is a random variable, the SNR is also random. In Fig.~\ref{fig:ICC14J_CRB}, we change the ToF measurement noise power (or equivalently the transmit power) to adjust the level of the ranging error at the average distance defined as $\sigma^\text{avg}_d = c\left(k_Ed^{n_0}_\text{avg}\right)^{\frac{1}{2}}$. 
As it can be anticipated from the theory of CRB, for low ToF noise power, the RMSE approaches its CRB, while for high noise power, it deviates from the CRB. 

In order to compare the total average energy consumption of the two schemes, the transmit and listening power values are selected from an actual underwater acoustic modem, the Evologics S2CR 12/24 \cite{Evologics12_24} as shown in Table~\ref{tab:env}. Using equations \eqref{eq:energy_CF} and \eqref{eq:energy_CT}, the average energy consumed by CFS and CTS is 30.14w and 12.72w, respectively. This indicates the higher energy consumption by CTS.
\section{Conclusion}
\label{sec:conclusion}
We have considered two classes of packet scheduling for self-localization in an underwater acoustic sensor network, one based on a collision-free design and another based on a collision-tolerant design. In collision-free packet scheduling, the time of the packet transmission from each anchor is set in such a way that none of the sensor nodes experiences a collision. In contrast, collision-tolerant algorithms are designed so as to control the probability of collision to ensure successful localization with a pre-specified reliability. We have also proposed a simple Gauss-Newton based localization algorithm for these schemes, and derived their Cram\'er-Rao lower bounds. The performance of the two classes of algorithms was shown to be comparable in terms of the time required for localization. When the ratio of the packet length to the maximum propagation delay is very low, the collision-tolerant protocol requires less time for localization in comparison with the collision-free one for the same probability of successful localization.
Other than average energy consumption by anchors, the collision-free scheme has other advantages with the major advantage being its simplicity of implementation with no requirements on synchronization. For a practical (non-zero) packet loss rate, collision-tolerant scheduling takes less time to localize a sensor node. In addition, the anchors work independently of each other, and as a result the scheme is spatially scalable, with no need for a fusion center. Finally, its localization accuracy is always better than that of the collision-tolerant scheme due to the multiple receptions of the desired packets from anchors. These features make the collision-tolerant localization scheme appealing for a practical implementation.   
In the future, we will extend this work to a multi-hop network where the communication range of the acoustic modems is much shorter than the size of the operating area.

\appendices
\section{$P^\text{loc}_{_\text{CT}}$ is an increasing function of $T_{_\text{CT}}$}
\label{app:p_loc_T}
In this appendix, we show that the probability of successful localization is an increasing function of the localization time. According to \eqref{eq:collisionTolPg}, and the fact that $p_s\lambda$ is independent of $T_\text{T}$, it is clear that $p_{_\text{CT}}$ is an increasing function of $T_\text{T}$. Therefore, $P^\text{loc}_{_\text{CT}}$ is an increasing function of $T_\text{T}$, if $P^\text{loc}_{_\text{CT}}$ is an increasing function of $p_{_\text{CT}}$. The derivative of $P^\text{loc}_{_\text{CT}}$ w.r.t. the $p_{_\text{CT}}$ is 
\begin{equation}
\footnotesize
\label{eq:app01_01}
\frac{\partial P^\text{loc}_{_\text{CT}}}{\partial p_{_\text{CT}}} = \sum_{k=K}^{N}{\left(\begin{matrix} N \\ k \end{matrix}\right)(k-Np_{_\text{CT}})p_{_\text{CT}}^{k-1}(1-p_{_\text{CT}})^{N-k-1}}.
\end{equation}
With a simple modification we have
\begin{equation}
\footnotesize
\hspace{-0.3cm}
\label{eq:app01_02}
\begin{aligned}
&\frac{\partial P^\text{loc}_{_\text{CT}}}{\partial p_{_\text{CT}}} = \frac{1}{p_{_\text{CT}}(1-p_{_\text{CT}})}\left\{ \begin{matrix}
 ~_{~}\\ ~_{~} 
\end{matrix}\right.\\
&\left[\sum_{k=0}^{N}{\left(\begin{matrix} N \\ k \end{matrix}\right)kp_{_\text{CT}}^{k}(1-p_{_\text{CT}})^{N-k}}\right. - 
\left.\sum_{k=0}^{K-1}{\left(\begin{matrix} N \\ k \end{matrix}\right)kp_{_\text{CT}}^{k}(1-p_{_\text{CT}})^{N-k}}\right] - \\
Np_{_\text{CT}}&\left[\sum_{k=0}^{N}{\left(\begin{matrix} N \\ k \end{matrix}\right)p_{_\text{CT}}^{k}(1-p_{_\text{CT}})^{N-k}}\right. - 
\left.\left.\sum_{k=0}^{K-1}{\left(\begin{matrix} N \\ k \end{matrix}\right)p_{_\text{CT}}^{k}(1-p_{_\text{CT}})^{N-k}}\right]\right\}.
\end{aligned}
\hspace{+0.4cm}
\end{equation}
Using the properties of binomial random variables we have that
\begin{equation}
\footnotesize
\label{eq:app01_03}
\sum_{k=0}^{N}{\left(\begin{matrix} N \\ k \end{matrix}\right)kp_{_\text{CT}}^{k}(1-p_{_\text{CT}})^{N-k}} = Np_{_\text{CT}},
\end{equation}
and
\begin{equation}
\footnotesize
\label{eq:app01_04}
\sum_{k=0}^{N}{\left(\begin{matrix} N \\ k \end{matrix}\right)p_{_\text{CT}}^{k}(1-p_{_\text{CT}})^{N-k}} = 1.
\end{equation}
Now, equation \eqref{eq:app01_02} (or equivalently \eqref{eq:app01_01}) is equal to
\begin{equation}
\footnotesize
\label{eq:app01_05}
\frac{\partial P^\text{loc}_{_\text{CT}}}{\partial p_{_\text{CT}}} = \sum_{k=0}^{K-1}{\left(\begin{matrix} N \\ k \end{matrix}\right)(Np_{_\text{CT}}-k)p_{_\text{CT}}^{k-1}(1-p_{_\text{CT}})^{N-k-1}}.
\end{equation}

It can be observed that \eqref{eq:app01_01} is always positive for $p_{_\text{CT}} < \frac{K}{N}$, and \eqref{eq:app01_05} is always positive for $p_{_\text{CT}} > \frac{K}{N}$, and as a result $\frac{\partial P^\text{loc}_{_\text{CT}}}{\partial p_{_\text{CT}}}$ is positive for any value of $p_{_\text{CT}}$; consequently, $P^\text{loc}_{_\text{CT}}$ is an increasing function of $p_{_\text{CT}}$, and consequently $T_\text{T}$.

\section{Maximum value of $p_s\lambda$}
\label{app:ps_lambda}
The first and second derivatives of $p_s\lambda$ w.r.t. $\lambda$ can be obtained as
\begin{equation}
\footnotesize
\label{eq:app02_01}
\frac{\partial p_s\lambda}{\partial \lambda} = \sum_{q=0}^{N}{p_{s|q}\frac{x^{q}e^{-x}}{q!}(q-x+1)},
\end{equation}

\begin{equation}
\footnotesize
\label{eq:app02_02}
\frac{(\partial p_s\lambda)^2}{\partial^2 \lambda} = \sum_{q=0}^{N}{p_{s|q}\frac{x^{q-1}e^{-x}}{q!}\left[(q-x)(q-x+1)-x\right]},
\end{equation}
where $x=2N\lambda T_p$. 
It can be observed that for $x<1$ the derivative in \eqref{eq:app02_01} is positive, and for $x>N+1$ it is negative. Therefore, $p_s\lambda$ has at least one maximum within $x\in[1,N+1]$. 
In practical scenarios the value of  $p_{s|q}$ for $k>0$ is usually small, so that it can be approximated by zero. For a special case where $p_{s|q>0} = 0$, \eqref{eq:app02_01} is zero if $x=1$, and \eqref{eq:app02_02} is negative, and as a result $\lambda^\text{low}_\text{opt}=\frac{1}{2NTp}$ maximizes $P^\text{loc}_{_\text{CT}}$. This corresponds to a lower bound on the optimal point in a general problem (i.e., $p_{s|q>0} \neq 0$).

\section{Cram\'er Rao lower bound for CTS}
\label{app:CRB_CTS}
As stated before, the upper bound on the sum operation in \eqref{eq:FIMeq} for CTS is $\infty$, and this makes the CRB calculation very difficult even if it is implemented numerically. In order to reduce the complexity of the problem, the observation of a sensor node from the $j$-th anchor is divided into two parts: Either a sensor node does not receive any packet from this anchor (no information is obtained), or it receives one or more packets. Since the anchor and the sensor node do not move very much during the localization procedure, their distance can be assumed almost constant, and therefore the noise power is the same for all measurements obtained from an anchor. When a sensor node gathers multiple measurements contaminated with independent noise with the same power (diagonal covariance matrix), the calculation of the CRB can be done with less complexity. We will explain this complexity reduction for the first anchor, and then we generalize this idea for the other anchors too. Considering the first anchor, each element of the FIM can be calculated in two parts; no correct packet reception from the first anchor, and one or more than one correct packet reception from this anchor which can be formulated as  
\begin{equation}
\footnotesize
{\bf I}({\bf x})_{i,j} = P^{0}{\bf I}({\bf x}|q_1 = 0)_{i,j} + P^{>0}{\bf I}({\bf x}|q_1 > 0)_{i,j}, \label{eq:FIManchor1}
\end{equation}
where $P^0=P^0_1$ is the probability that no packet is received from an anchor which is same for all anchors, and  $P^{>0}=P^{>0}_1=\sum_{q_1=1}^{\infty}{P_1^k}$ is the probability that one or more than one packets are received from an anchor which is also same for all anchors. The second term in \eqref{eq:FIManchor1} can be expanded as
\begin{equation}
\footnotesize
\begin{aligned}
{\bf I}&({\bf x}|q_1 > 0)_{i,j} = \frac{1}{P^\text{loc}} \underset{\text{s.t. } \{q_1,...,q_N\} \text{ enable self-localization}}{\sum_{{q_N=0}}^{Q_N} \hdots\sum_{q_2 = 0}^{Q_2}} \\
&\left\{1\sigma_1^{-2}\frac{\partial f_1}{\partial x_i}\frac{\partial f_1}{\partial x_j} + c_1 + 1\sigma_1^{-4}\frac{\partial \sigma_1^{2}}{\partial x_i}\frac{\partial \sigma_1^{2}}{\partial x_j}+c_2\right\} {P_1^1}/{P_1^{>0}}\Pi_{j=2}^{N}P^{q_j}_{j} + \\
&\left\{2\sigma_1^{-2}\frac{\partial f_1}{\partial x_i}\frac{\partial f_1}{\partial x_j} + c_1 + 2\sigma_1^{-4}\frac{\partial \sigma_1^{2}}{\partial x_i}\frac{\partial \sigma_1^{2}}{\partial x_j}+c_2\right\} {P_1^2}/{P_1^{>0}}\Pi_{j=2}^{N}P^{q_j}_{j} + \\
&~~~~~~~~~~~~~~~~~~~\vdots \\
&\left\{k\sigma_1^{-2}\frac{\partial f_1}{\partial x_i}\frac{\partial f_1}{\partial x_j} + c_1 + k\sigma_1^{-4}\frac{\partial \sigma_1^{2}}{\partial x_i}\frac{\partial \sigma_1^{2}}{\partial x_j}+c_2\right\} {P_1^k}/{P_1^{>0}}\Pi_{j=2}^{N}P^{q_j}_{j} + \\
&~~~~~~~~~~~~~~~~~~~\vdots \\
\end{aligned}
\end{equation}
where $c_1$ and $c_2$ are only affected by measurements from the other anchors. Using a simple factorization we have  
\begin{equation}
\footnotesize
\begin{aligned}
{\bf I}&({\bf x}|q_1 > 0)_{i,j} = \frac{1}{P^\text{loc}} \underset{\text{s.t. } \{q_1,...,q_N\} \text{ enable self-localization}}{\sum_{{q_N=0}}^{Q_N} \hdots\sum_{q_2 = 0}^{Q_2}} \\
&\left\{ g_{_\text{CT}}\left[\sigma_1^{-2}\frac{\partial f_1}{\partial x_i}\frac{\partial f_1}{\partial x_j} + \sigma_1^{-4}\frac{\partial \sigma_1^{2}}{\partial x_i}\frac{\partial \sigma_1^{2}}{\partial x_j}\right] + c_1 +c_2\right\}\Pi_{j=2}^{N}P^{q_j}_{j}
\end{aligned}
\label{eq:simplifiedFIM}
\end{equation}
where 
\begin{equation}
\footnotesize
g_{_\text{CT}}=\frac{\sum_{q_1=1}^{\infty} kP_1^k}{\sum_{q_1=1}^{\infty} P_1^k} = \frac{ p_s\lambda T_\text{T}}{1- P^0}
\end{equation}
can be calculated analytically and be used in \eqref{eq:simplifiedFIM}. That enables us to calculate only two possible states for the sum over $q_1$. Now, we define ${\bf a}_{N \times 1}$ where its $k$-th element $a_k$ is either zero (if $q_k=0$) or 1 (if $q_k > 0$). We also define ${\bf b}_{N \times 1}$ with its $k$-th element $b_k = \left[\sigma_k^{-2}\frac{\partial f_k}{\partial x_i}\frac{\partial f_k}{\partial x_j} + \sigma_k^{-4}\frac{\partial \sigma_k^{2}}{\partial x_i}\frac{\partial \sigma_k^{2}}{\partial x_j}\right]$. Then, we have
\begin{equation}
\footnotesize
{\bf I}({\bf x}|{\bf a})_{i,j} = {g_{_\text{CT}}}\frac{1}{P^\text{loc}}{\bf a}^T{\bf b}\left(P^0\right)^{N-n_{\bf a}}\left(1-P^0\right)^{n_{\bf a}}. 
\end{equation} 
where $n_{\bf a}$ is the number of non-zero elements in ${\bf a}$. This means that to evaluate ${\bf I}({\bf x})_{i,j}$ for the localizable scenarios only $\footnotesize
{\left(\begin{matrix}
N\\K
\end{matrix}\right)}$ possible states (different realizations of ${\bf a}$ which lead to localizable scenarios) have to be considered which is the same as that of CFS.

\section{distribution of the mutual distance}
\label{app:pdf_d}
\begin{figure}[h]
\centering
\includegraphics[width=.30\textwidth]{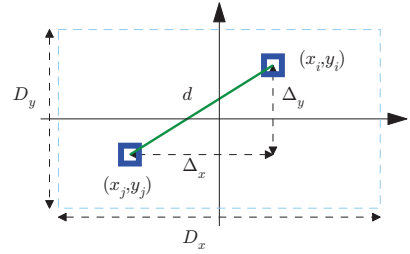}
\caption{\footnotesize{Two randomly located nodes in a rectangular operating area.}}
\label{fig:LMAC_FIG_Pdf01}
\end{figure}

\begin{figure}[h]
\centering
\vspace{-0.4cm}
\includegraphics[width=.40\textwidth]{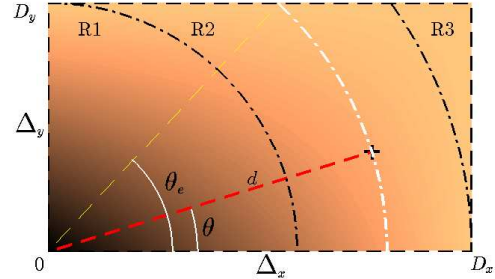}
\caption{\footnotesize{Illustration of the parameters and their relations to each other in calculating the pdf of the distance between two nodes located uniformly at random. }}
\label{fig:LMAC_FIG_Pdf02}
\end{figure}

In this appendix, we derive the pdf of the distance between two nodes located uniformly at random in a rectangular region as shown in Fig.~\ref{fig:LMAC_FIG_Pdf01}. 
Under this condition the pdfs of the $x$ and $y$ projections of the distance are
\begin{subequations}
\footnotesize
\begin{eqnarray}
f_{\Delta_X}(\Delta_x)=\frac{2}{D_x^2}(D_x-\Delta_x),~ 0 \leq \Delta_x \leq D_x\\
f_{\Delta_Y}(\Delta_y)=\frac{2}{D_y^2}(D_y-\Delta_y),~ 0 \leq \Delta_y \leq D_y,
\end{eqnarray}
\end{subequations}
and since they are independent, the joint pdf in polar coordinates (see Fig.~\ref{fig:LMAC_FIG_Pdf02}) is
\begin{equation}
\footnotesize
f_{D,\Theta}(d,\theta)=\frac{4d}{D_x^2D_y^2}(D_x-d\cos\theta)(D_y - d\sin\theta).
\end{equation}

\begin{figure}[h]
\centering
\includegraphics[width=.35\textwidth]{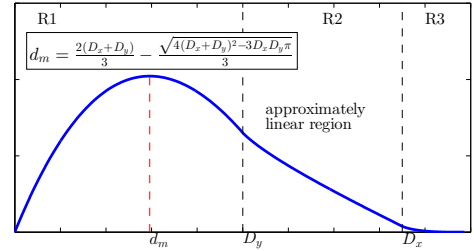}
\caption{\footnotesize{Probability density function of the distance between two uniformly randomly located nodes. $d_m$ is the point at which the maximum of the pdf occurs.}}
\label{fig:LMAC_FIG_Pdf03}
\end{figure}

By taking an integral over $\theta$, the pdf of the distance follows \eqref{eq:distPdf}. This pdf is shown in Fig.~\ref{fig:LMAC_FIG_Pdf03}.

\bibliographystyle{IEEEtran}
\small
\bibliography{ICC14_bib}

\end{document}